\pdfoutput=1
\documentclass[aps,prl,twocolumn,10pt,superscriptaddress,longbibliography]{revtex4-2}

\usepackage{graphicx}   
\usepackage{CJKutf8}
\usepackage{amsmath,amssymb}
\usepackage{bm}         
\usepackage{siunitx}    
\usepackage[colorlinks=true,allcolors=blue]{hyperref}
\usepackage{microtype}
\usepackage{xcolor}

\begin{document}
\begin{CJK*}{UTF8}{gbsn}

\title{Attraction kinetics controls permanent shear banding in particulate dispersions}

\author{Liheng Yao (姚立衡)}
\author{Romain Mari}
\author{Kirsten Martens}
\email{}
\affiliation{Université Grenoble Alpes, CNRS, LIPhy, 38000 Grenoble, France}

\begin{abstract}
We investigate the connection between kinetic attractive interactions (i.e. particle attraction that turn on and off at finite rates) and the formation of permanent shear bands in particulate systems. To that end, we develop a mean field constitutive theory based on the idea that, under shear, kinetic attraction generates an overstress only when the shear rate is comparable to the attraction kinetic rates. We verify our theory using simulations of a jammed particle system with kinetic attractive interactions. We measure the homogeneous flow curves of our system at small system sizes, and find them to be in good qualitative agreement with our theory. At larger system sizes, permanent shear banding is observed. We discuss the generality of our theoretical framework.
\end{abstract}

\maketitle
\end{CJK*}

\paragraph*{Introduction --}

Permanent shear banding, i.e., the stable coexistence of regions flowing at different (possibly vanishing) shear rates under a common shear stress, is observed in a variety of complex fluids, including wormlike micelles, colloidal suspensions, emulsions, and granular materials \cite{olmstedPerspectivesShearBanding2008,coussotPhysicalOriginShearbanding2010,fieldingShearBandingSoft2014,divouxShearBandingComplex2016}. Permanent shear banding is often associated with a non-monotonic constitutive relation, where the stress decreases with increasing shear rate, rendering homogeneous flow unstable. Several microscopic mechanisms have been identified, including nonlinear viscoelasticity and flow-induced alignment in wormlike micelles \cite{spenleyNonlinearRheology1993}, shear-concentration coupling in colloidal glasses \cite{besselingShearBanding2010}, shear-induced weakening of frictional contacts \cite{degiuliFrictionLaw2017} and contact aging \cite{jaglaShearBandsEarthquakes2024} in granular materials, local inertial effects in amorphous solids \cite{nicolasEffectsInertia2016,vasishtPermanentShearLocalization2023}, and slow structural recovery following plastic rearrangements \cite{coussotPhysicalOriginShearbanding2010,martensSpontaneousFormation2012}.

Attractive interactions influence yielding and flow heterogeneity in a range of particulate systems, including adhesive emulsions \cite{becuYieldingFlowAdhesive2006}, colloidal gels \cite{gibaudHeterogeneousYielding2010,libertoInterparticleAttraction2020}, and adhesive granular gels \cite{chattopadhyayEffectAdhesiveInteraction2022}. Related phenomena have been reported in aging soft-colloidal suspensions and thixotropic materials \cite{rogersAgingYieldingShear2008,coussotViscosityBifurcationThixotropic2002,divouxRheologicalHysteresisSoft2013}. However, attraction alone does not necessarily produce permanent shear banding. Numerical studies show that static attractive interactions can induce permanent shear banding below but close to the jamming transition \cite{iraniImpactAttractiveInteractions2014}, whereas above jamming they fail to produce permanent shear bands \cite{chaudhuriInhomogeneousShearFlows2012, Yankah2026Sep}.

\begin{figure}
\centering
\includegraphics[width=1.0\columnwidth]{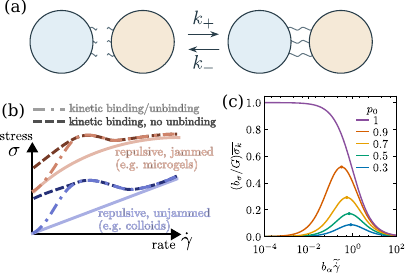}
\caption{(a) An illustration of a particle pair experiencing kinetic attraction. (b) Schematic flow curves of particulate systems with and without kinetic attraction. (c) Plot of the excess stress due to the kinetic attraction as given in \eqref{eq:steady_flow_curve}, for various values of $p_0$, assuming $b_{\alpha} / b_{\sigma} = 1$.}
\label{fig:fig1}
\end{figure}

However, in dense particulate dispersions, attractive interactions arise from a variety of microscopic mechanisms \cite{Israelachvili2011,likosEffectiveInteractionsSoft2001}, often involving finite association and dissociation timescales (see Fig.~\ref{fig:fig1}(a)). Examples include depletion-induced interactions in binary colloids \cite{crockerEntropicAttractionRepulsion1999}, attraction across electrostatic repulsive barriers \cite{moazzami-gudarziDepletionDoubleLayer2017}, ligand-mediated and DNA-mediated adhesion \cite{leunissenSwitchableSelfprotectedAttractions2009,rogersKineticsNonexponentialBinding2013,wuKineticsDNAcoatedSticky2013,cuiComprehensiveViewMicroscopic2022}. Such interactions are also present in colloidal self-assembly \cite{stoevBulkRheologySticky2021,jacobsAssemblyComplexColloidal2025}, patchy colloids \cite{wangColloidsValenceSpecific2012,kraftSurfaceRoughnessDirected2012}, and clay suspensions \cite{sahaDynamicLightScattering2015}. Mattiello and Cloître observed persistent shear banding in concentrated suspensions of microgels with short-range attractive interactions governed by reversible bond association and dissociation \cite{ Mattiello2018}. When the characteristic timescales of these interactions become comparable to that of deformation, the mechanical response may depend not only on the strength of attraction but also on the rates at which attractive bonds form and break. The connection between such kinetics and permanent shear banding in jammed particulate systems remains an open question.

In this Letter, we address the link between attraction kinetics and permanent shear banding in dispersions, introducing a minimal rheological model for the overstress due to the presence of transient attractive bonds.
In this model the attractive bonds give a visco-elastic contribution to the total stress, with a characteristic bell shape as a function of the applied shear rate.
Combined with a base rheology, with or without yield stress depending on the system, this overstress can lead to non-monotonic flow curves under specific conditions for the binding and unbinding rates. 
We exemplify this behavior for a yield stress fluid with particle-based simulations of a system of jammed particles undergoing kinetic attraction, whose flow curves and phase diagram are in good agreement with the predictions of our model.
Beyond this example, our results highlight the key kinetic parameters and conditions leading to a flow instability in particulate dispersions with kinetic attractive interactions.

\paragraph*{Theory --}
In particulate systems with very fast attractive binding, the density of attractive bonds is essentially controlled by the flow, which brings pairs of particles together and apart, thereby building up and destroying attractive bonds.
By contrast, when the binding/unbinding kinetic rates are comparable to the rate of deformation, the bond density under flow can be limited by the reaction kinetics. 
We sketch in Fig.~\ref{fig:fig1}(b) the anticipated effects on the rheology, which are qualitatively similar whether the equivalent system with no attractive forces at all is below jamming (blue-tinted flow curves) or above jamming with a yield stress (red-tinted flow curves). We write the binding rate as $k_+$ and the unbinding rate as $k_-$ (see Fig.~\ref{fig:fig1}(a)).
At large shear rates $\dot\gamma\gg k_+$, bonds have no time to form and the particles remain purely repulsive. At rates $\dot\gamma \lesssim k_+$, the typical time two neighboring particles spend within interaction distance is comparable to or longer than that of the typical binding, so that a finite density of attractive bonds form. 
In this case, the effect of attractive bonds on the rheology depends on $k_-$.
If  $\dot\gamma \gtrsim k_-$, attractive bonds remain active long enough to be stretched by the flow and resist it, giving rise to an extra stress contribution on top of the one provided by repulsive interactions.
On the contrary, if $\dot\gamma \ll k_-$, the lifetimes of the attractive bonds vanish on the timescale of nearest-neighbor persistence, making the attraction ineffective in building up this excess stress.
From this analysis, we therefore expect that for finite values of $k_+, k_-$ such that $0< k_{\pm} < \infty$, the contribution to the stress of the attractive bonds is a non-monotonic function of $\dot\gamma$, vanishing for small and large rates, but with a bell shape for $\dot\gamma$ in a range around $[k_-, k_+]$. 
This non-monotonic contribution may lead to a non-monotonic total stress, depending on the slope of the underlying flow curve, with the decreasing part of the flow curve lying in between two stable increasing branches at small and large rates $\dot\gamma$ (dot-dashed lines in Fig.~\ref{fig:fig1}(b)). 
For the special case where spontaneous unbinding is absent ($k_-=0$), attraction always plays a role at $\dot\gamma\to 0$ and may create a yield stress or increase its value compared to the equivalent purely repulsive system (dashed lines in Fig.~\ref{fig:fig1}(b)).

We introduce a mean-field constitutive relation for a particle system with kinetic attractive interactions that embodies these ideas. We describe the shear stress of the system as $\sigma = \sigma_0 + \sigma_k$, where $\sigma_0$ is the underlying constitutive law of the system in the absence of the attractive interactions, while $\sigma_k$ describes the contribution of the kinetic attraction. We describe the bond-induced stress as being loaded elastically by shear while relaxing through dissociation and deformation, such that
\begin{equation}
    \frac{\partial\sigma_k}{\partial t}
    = G\alpha\dot\gamma
    - (k_-+b_{\sigma}\dot\gamma)\sigma_k,
    \label{eq:bond_stress_dynamics}
\end{equation}
where $G$ is related to the elastic modulus of each bond, and can potentially be measured in the linear response regime~\cite{Shahmohammadi2021May}. The number of bonds in the system is described by the bond occupancy parameter $0 \leq \alpha \leq 1$, defined as the ratio between the number of bonds in the system and the number of particle pairs eligible to bond forming. $b_{\sigma}$ describes shear-induced stress loss due to the re-orientation, stretching or breaking of bonds. 

The dynamics of $\alpha$ is then controlled by the competition between bond formation and dissociation:
\begin{equation}
    \frac{\partial\alpha}{\partial t}
    = k_+(1-\alpha)
    - (k_-+b_{\alpha}\dot\gamma)\alpha,
    \label{eq:alpha_dynamics}
\end{equation}
where $k_+$ and $k_-$ are the kinetic association and dissociation rates, and $b_{\alpha}$ controls shear-induced bond breakage.

We define two control parameters: the equilibrium bond fraction $p_0 = k_+ / (k_+ + k_-)$ and the dimensionless shear rate $\widetilde{\dot\gamma} = \dot\gamma / \lambda_k$. Here $\lambda_k = k_+ + k_-$ gives the inverse timescale of the kinetic interactions. The steady-state excess stress due to the kinetic attraction is
\begin{equation}
\overline{\sigma_k} \left(p_0, \widetilde{\dot\gamma} \right)
= \frac{G p_0\widetilde{\dot\gamma}}
{\left(1+b_\alpha\widetilde{\dot\gamma}\right)
 \left(1-p_0+b_\sigma\widetilde{\dot\gamma}\right)} ,
\label{eq:steady_flow_curve}
\end{equation}
which is plotted in Fig.~\ref{fig:fig1}(c), assuming $b_{\alpha} / b_{\sigma} = 1$ without loss of generality (changing this ratio shifts the peaks of $\overline{\sigma_k}$ for all $p_0$). The competition between bond formation and breakage produces an effective rate-dependent damping term in the constitutive relation, which causes this excess stress to decrease with increasing shear rate in some regimes. For $p_0 < 1$, in the regime where $\widetilde{\dot\gamma}$ is very large or very small, $\overline{\sigma_k} \rightarrow 0$, so that the stress of the system converges to the stress of the underlying system, in agreement with the physical picture discussed earlier. At precisely $p_0 = 1$, which corresponds to a system without spontaneous bond breakage, the bond-carried excess stress converges to a positive constant $G / b_\sigma$ at low shear rates. In the even more stringent limit of a non-kinetic fully attractive interaction, where $p_0 = 1$ and $\lambda_k \rightarrow \infty$, the bond-carried stress becomes a constant at all shear rates, and the shape of the flow curve will mirror that of the underlying constitutive law $\sigma_0$. Therefore, if the underlying system does not exhibit flow instabilities, neither will the the system with added non-kinetic attraction. This is in agreement with known results for jammed and overdamped attractive systems, where non-kinetic attractive interactions do not induce permanent shear banding \cite{chaudhuriInhomogeneousShearFlows2012, Yankah2026Sep}.

\begin{figure}[tbp]
\centering
\includegraphics[width=\columnwidth]{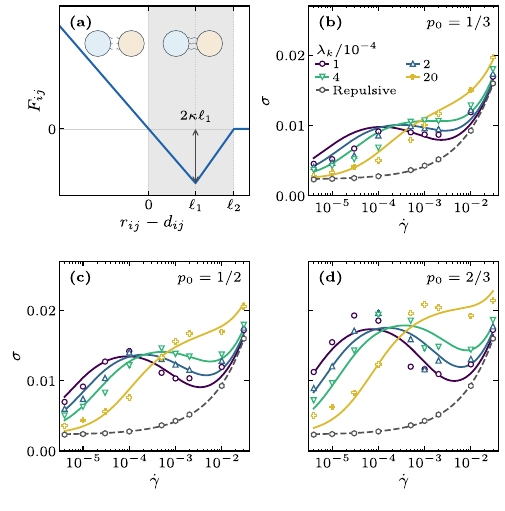}
\caption{(a) An illustration of the interaction force due to the potential \eqref{eq:int} in the particle model. The shaded part of the interaction is only present when $s_{ij} = 1$, i.e. when the particle pair is in the bonded state.  (b -- d) The intrinsic flow curves of the particle system at $p_0 = (1/3, 1/2, 2/3)$ and $\lambda_k / 10^{-4} = (1, 2, 4, 20)$, and the fully repulsive system. The symbols are the data points from the numerical simulation at system size $V = 10 \times 10 \times 5$, while the lines are fits of \eqref{eq:steady_flow_curve} at $b_{\alpha} = 0.092$, $b_{\sigma} = 9.6$ and $G = 0.24$, added to a Herschel-Bulkley law $\sigma_0 = \sigma_{Y} + K \dot \gamma^n$ with $\sigma_Y = 0.0023$, $K = 0.11$ and $n = 0.59$. Here, we use the simulation units of length, time, and energy, which are $\bar{d}$, $\zeta / 4 \kappa$, and $2 \kappa \bar{d}^2$, respectively.}
\label{fig:flow_curves_small}
\end{figure}

\paragraph*{Particle model -- }

To test our framework, we simulate an athermal suspension of overdamped soft particles in 3D using LAMMPS \cite{LAMMPS}. The system contains equal numbers of bidisperse particles, with average diameter $\bar{d} = 1$ and diameter ratio $1.4$. We set the number density to $\rho=1.24$, so that the system is well in the jammed regime even without attractive interactions.

Consider two particles $i$ and $j$ with radii $a_i$ and $a_j$, respectively. We denote their center-to-center separation by $r_{ij}$, define their contact distance as $d_{ij}=a_i+a_j$, and take the unit vector $\hat{\mathbf{r}}_{ij}$ to point from particle $j$ to particle $i$. The force on $i$ due to $j$ consists of repulsive and attractive parts, and corresponds to a potential that is piecewise harmonic:
\begin{equation}
\mathbf{F}_{ij}
=\left[f_{\mathrm{rep}}(r_{ij};d_{ij})
+s_{ij}f_{\mathrm{att}}(r_{ij};d_{ij})\right]\hat{\mathbf{r}}_{ij},
\label{eq:int}
\end{equation}
where $s_{ij}=0$ for an unbonded pair and $s_{ij}=1$ for a bonded pair. The repulsive part of the force is $f_{\mathrm{rep}}(r_{ij};d_{ij})=2\kappa(d_{ij}-r_{ij})$ for $r_{ij}<d_{ij}$ and zero otherwise. We set $\kappa=0.5$.

The attractive part of the force has an outer cutoff $r_{c,ij}=d_{ij}+\ell_2$. Within this range, unbonded pairs associate at rate $k_+$ and bonded pairs dissociate at rate $k_-$. The attractive force is given by
\begin{equation}
\begin{gathered}
f_{\mathrm{att}}(r_{ij};d_{ij})=\\[-2pt]
\begin{cases}
0, & r_{ij}<d_{ij},\\
-2\kappa(r_{ij}-d_{ij}),
    & d_{ij}\le r_{ij}\le d_{ij}+\ell_1,\\
-\displaystyle\frac{2\kappa\ell_1}{\ell_2-\ell_1}(r_{c,ij}-r_{ij}),
    & d_{ij}+\ell_1<r_{ij}<r_{c,ij},\\
0, & r_{ij}\ge r_{c,ij}.
\end{cases}
\end{gathered}
\label{eq:att}
\end{equation}
We show a schematic plot of the interaction in Fig.~\ref{fig:flow_curves_small}(a). In the non-kinetic limit, this interaction is similar to that used in \cite{chaudhuriInhomogeneousShearFlows2012, iraniImpactAttractiveInteractions2014}. For this work, we choose $\ell_1=0.099$ and $\ell_2=0.1$, which corresponds to a bond length that is 10\% of the average particle diameter. Bonds also break deterministically when $r_{ij}$ exceeds $r_{c,ij}$.

Dissipation in our model is implemented through dissipative particle dynamics (DPD)~\cite{Hoogerbrugge1992Jun}, where neighbouring particle pairs experience a momentum-conserving radial drag $\mathbf F^{\mathrm{DPD}}_{ij}=-\zeta\big[(\mathbf v_i-\mathbf v_j)\cdot\hat{\mathbf r}_{ij}\big]\hat{\mathbf r}_{ij}$ that vanishes for $r_{ij}\ge d_{ij}$. Newton's equations are integrated using the velocity Verlet algorithm using a timestep $\Delta t=0.02$, with all particle masses set to 1. To simulate an overdamped system, we set $\zeta=2$. In the following simulations, we use the average particle diameter $\bar{d}$, the dissipative relaxation time $\zeta / 4 \kappa$, and the elastic energy scale $2 \kappa \bar{d}^2$ as units of length, time, and energy, respectively.

\begin{figure}[tbp]
\centering
\includegraphics[width=\columnwidth]{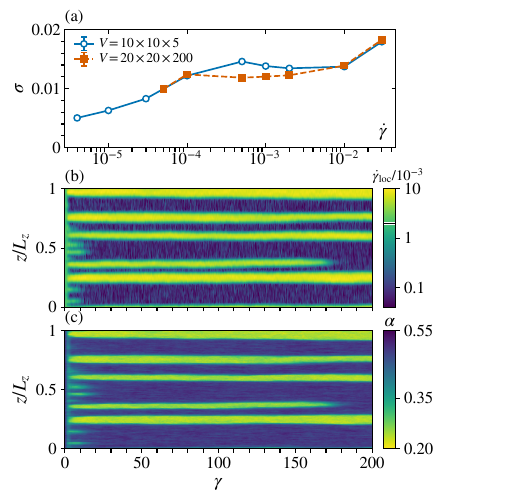}
\caption{(a) The flow curves of a $p_0 = 1/2$, $\lambda_k = 4 \times 10^{-4}$ system at different system sizes. (b, c) Space-time plots of the local shear rate and the local bond occupancy for this system at size $V = 20 \times 20 \times 200$, driven at $\dot \gamma = 2 \times 10^{-3}$.}
\label{fig:spacetime}
\end{figure}

\paragraph*{Numerical results --} To investigate the intrinsic flow curves of our particle model in its homogeneous state, we perform simple shear trajectories on small systems of size $L_x = L_y = 10$, $L_z = 5$, with $z$ being the gradient direction (the small system size suppresses shear banding). We prepare a well-relaxed homogeneous initial state for each choice of parameters, and shear well into the steady state using total strains between 18 and 70, such that all data points in Fig.~\ref{fig:flow_curves_small}(b -- d) are averaged over 10 steady-state strain units. 
We group our flow curves according to the static steady-state bond fraction $p_0$. The results show non-monotonic flow behavior. In particular, the individual kinetic rates control the onset of the unstable region, such that varying $\lambda_k$ shifts the flow curve even for systems with the same $p_0$. For comparison, we also plot the flow curve for the system with a purely repulsive interaction potential $f_{\mathrm{rep}}$.

We can now produce a fit of our mean field theory to the data. The underlying constitutive relation of this system in the absence of kinetic attraction is well described by the Herschel-Bulkley law $\sigma_0 = \sigma_{Y} + K \dot \gamma^n$, with parameters that can be fitted to the flow curve of the fully repulsive system. In our simulations we can also measure the parameter $\alpha$ by measuring the number of bonds in the system and the total number of eligible bond-forming pairs. This allows us to perform a fit of $b_\alpha$ independent of the stress measurements (see Fig.~\ref{fig:alpha} in the Supplementary Materials (SM)). We can now fit all of our stress data from the kinetically attractive systems to two fit parameters $G$ and $b_\sigma$. The fitted curves are shown in Fig.~\ref{fig:flow_curves_small}(b -- d). This fit, while short of a precise quantitative description of our data, correctly describes the dependence of the flow curves on the kinetic parameters $p_0$ and $\lambda_k$.

We now demonstrate permanent shear banding in our model by moving to a large system with $L_x = L_y = 20$ and $L_z = 200$. We focus on a system with $p_0 = 1/2$, $\lambda_k = 4 \times 10^{-4}$ sheared at $\dot \gamma = 2 \times 10^{-3}$, corresponding to a point on the negative branch of the intrinsic flow curve. At this system size the flow curve shows significantly less non-monotonicity, as seen in Fig.~\ref{fig:spacetime}(a) in agreement with experimental findings (see Fig.~5.15 in Ref.~\cite{Mattiello2018}). For a long trajectory starting from the stationary steady state, the spacetime plot of the local shear rate in Fig.~\ref{fig:spacetime}(b) shows the formation of shear bands followed by slow coarsening. At large strains, we expect the system to eventually coarsen to contain a single shear band, as shown in Fig.~\ref{fig:sb} in the SM, where the same system as in Fig.~\ref{fig:spacetime} seeded with a single shear band retains that shear band indefinitely.
Another spacetime plot of $\alpha$ in Fig.~\ref{fig:spacetime}(c) shows that the spatially resolved evolution of the local shear rate is strongly synchronous with that of the bond fraction $\alpha$. This confirms that the bond-forming kinetics is indeed the controlling mechanism of the shear-banding instability. We note that, in experiments, typically only a single shear band is observed, often localized near the system boundary due to pronounced boundary-induced heterogeneities (see Fig.~5.17 in Ref.~\cite{Mattiello2018}).

\begin{figure}[tbp]
\centering
\includegraphics[width=0.65\columnwidth]{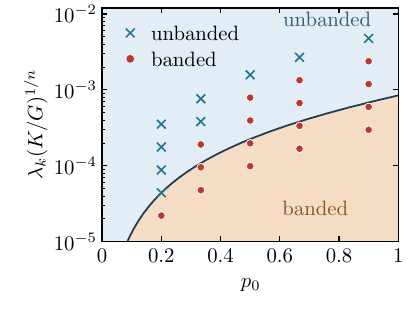}
\caption{The phase diagram of the particle system. The data points are produced using a system of size $V = 10 \times 10 \times 100$, while the solid line is the spinodal predicted by the theory with the same fitting parameters as Fig.~\ref{fig:flow_curves_small}.}
\label{fig:phase}
\end{figure}

We can also produce a phase diagram of our particle system, as shown in Fig.~\ref{fig:phase}, in the plane of dimensionless rate $\lambda_k(K/G)^{1/n}$ and equilibrium bond fraction $p_0$.
We perform linear stability analysis (see SM) of the theory using the fitting parameters from the homogeneous flow curves in Fig.~\ref{fig:flow_curves_small}(b -- d) to estimate the boundary of linear stability between ``banded'' systems, where homogeneous flow is linearly unstable for some shear rates, and ``unbanded'' systems, where homogeneous flow is stable for all shear rates. The resultant spinodal curve is the solid line in Fig.~\ref{fig:phase}. 
This is then compared against explicit simulations at system size $L_x = L_y = 10$ and $L_z = 100$, indicated by the data points in Fig.~\ref{fig:phase}.
While the binodal separating banded and unbanded phases does not quantitatively track the mean-field model spinodal, the general trend in the dependence of the flow homogeneity on the kinetic rates is still well captured by our simple mean field description.
In particular, banding occurs at kinetic rates $\lambda_k(K/G)^{1/n}$ below a critical value which increases with increasing equilibrium bond fraction $p_0$. In the limit of infinitely fast dissociation $p_0\to 0$, attractive bonds are too short-lived to induce any significant stress and shear banding does not occur.

\paragraph*{Discussion --}
In this Letter, we have developed a general mean-field constitutive theory for particulate systems with kinetic attractive interactions. We show that the competition between the reaction and shear timescales can induce a non-monotonic flow curve, which for large systems lead to a shear banding instability. Numerical simulations of a system of jammed soft particles with kinetic attraction are in good qualitative agreement with our theory. Our theory is also able to explain the lack of permanent shear banding in similar systems with non-kinetic attractive interactions. Since our physical arguments are independent of the underlying constitutive laws of the repulsive system, we anticipate that our framework will be able to explain the scaling of the flow instability with the kinetic timescales for general kinetically attractive particulate systems.

Our theory neglects spatial gradient terms to retain a general description independent of the microscopic details of a particular system. Including such terms would allow a more detailed analysis of the instability growth rates and stress selection in the shear-banded state, but their form is expected to be system-specific. Furthermore, attractive interactions induce significant particle clustering in our simulations, causing the assumption of a bond rupture rate linear in shear rate to become less accurate at high shear rates. Accounting for nonlinear bond rupture kinetics could improve quantitative agreement with simulations and experiments, although the appropriate functional dependence would likewise depend on the system considered.

In many engineering applications it is important to control shear instabilities to mainly avoid them. The phase diagram in Fig.~\ref{fig:phase} identifies several possibilities to suppress shear banding: reducing the equilibrium bond fraction $p_0$, increasing the overall kinetic rate $\lambda_k$, or increasing the ratio $K/G$ between the repulsive and attractive stress scales. In particular, increasing $\lambda_k$ at fixed $p_0$ can restore homogeneous flow without changing the equilibrium fraction of attractive bonds. These parameters could be tuned through changes in temperature, particle interactions, or the properties of the suspending medium, depending on the system. Our results thus suggest different strategies to suppress persistent flow heterogeneity, either by modifying attraction kinetics or by changing the relative contributions of repulsive and attractive interactions to the rheological response.

\begin{acknowledgments}
We thank Michel Cloître, Paolo Edera, Stefano Aime, Pinaki Chaudhuri and Vishwas Vasisht for fruitful discussions. This work was supported by the
French Agence Nationale de la Recherche (ANR), grant No. ANR-24-CE06-1565
(HistoRYS project). Numerical simulations have been performed
on the GRICAD infrastructure (https://gricad.univ-grenoble-alpes.fr), which is supported by the Grenoble
research communities.
\end{acknowledgments}

\bibliographystyle{apsrev4-2}
\bibliography{refs} 

\clearpage
\section*{Supplementary Materials}

\subsection{Supplementary figures}

\begin{figure}
\centering
\includegraphics[width=0.65\columnwidth]{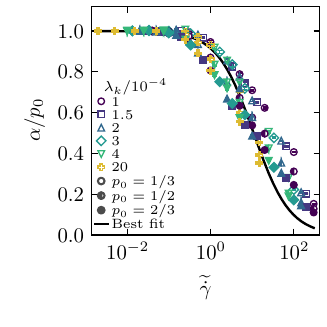}
\caption{Data collapse of the bond occupancy parameter $\alpha$ from our particle simulations. The fitted curve gives $b_{\alpha} = 0.092$.}
\label{fig:alpha}
\end{figure}

\begin{figure}
\centering
\includegraphics[width=\columnwidth]{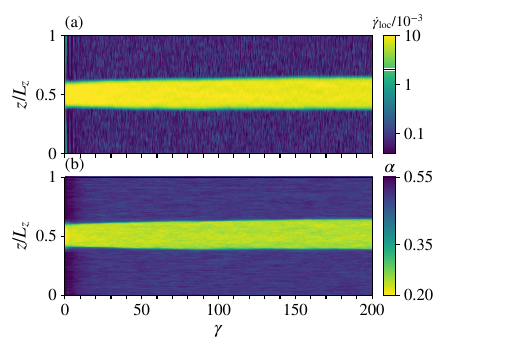}
\caption{The spacetime plot of (a) the local shear rate and (b) the bond occupancy of a  $p_0 = 1/2$, $\lambda_k = 4 \times 10^{-4}$ system of size $V = 20 \times 20 \times 200$, with initial conditions that seed a single shear band into the system.}
\label{fig:sb}
\end{figure}

Fig.~\ref{fig:alpha}: Data collapse of the bond occupancy parameter $\alpha$ from our particle simulations, from which a fit of $b_{\alpha}$ can be performed.

Fig.~\ref{fig:sb}: The spacetime plots of the local shear rate and bond occupancy of a system with the same set of parameters as in Fig.~\ref{fig:spacetime}(b, c) ($p_0 = 1/2$, $\lambda_k = 4 \times 10^{-4}$, $\dot \gamma = 2 \times 10^{-3}$ and $V = 20 \times 20 \times 200$), but with initial conditions that seed a single shear band into the system. To implement this seeding, we start from a sample in its stationary steady state, and apply velocity and bond occupancy perturbations following a smooth function
\begin{equation}
W(z)=\frac{1}{2}\left[
\tanh\!\left(\frac{z-80}{5}\right)
-\tanh\!\left(\frac{z-120}{5}\right)
\right].
\end{equation}
The strength of the perturbations are chosen to preserve the overall average bond occupancy, and impose a global shear rate equal to the applied $\dot \gamma = 2 \times 10^{-3}$. The system is then allowed to evolve under its natural dynamics.
The figure shows that the seeded shear band survives indefinitely, indicating that the true steady state for these parameters involves a single permanent shear band.

\subsection{Linear stability of homogeneous shear}
\label{sec:linear_stability}

We consider positive shear rates, without loss of generality. Since the underlying constitutive law of the particle model is Herschel-Bulkley, we have the local constitutive equations
\begin{align}
    \partial_t\alpha
    &=
    k_+(1-\alpha)
    -(k_-+b_\alpha\dot\gamma)\alpha,
    \label{eq:structure_local}
    \\
    \partial_t\sigma_k
    &=
    G\alpha\dot\gamma
    -(k_-+b_\sigma\dot\gamma)\sigma_k,
    \label{eq:stress_local}
    \\
    \sigma
    &=
    \sigma_Y+K\dot\gamma^n+\sigma_k.
    \label{eq:total_stress_local}
\end{align}
Here, the structural and mechanical
variables are coupled through force balance and the
constraint that the mean shear rate be equal to the imposed value.

\paragraph{Homogeneous steady state.}

At an imposed mean shear rate $\overline{\dot\gamma}>0$, define
\begin{equation}
    A=k_++k_-+b_\alpha\overline{\dot\gamma},
    \qquad
    B=k_-+b_\sigma\overline{\dot\gamma}.
\end{equation}
The homogeneous steady state is
\begin{equation}
    \overline\alpha=\frac{k_+}{A},
    \qquad
    \overline{\sigma_k}=\frac{G\overline\alpha\overline{\dot\gamma}}{B},
\end{equation}
with total stress
\begin{equation}
    \overline\sigma(\overline{\dot\gamma})
    =
    \sigma_Y+K\overline{\dot\gamma}^n
    +\frac{Gk_+\overline{\dot\gamma}}{AB}.
    \label{eq:homogeneous_flow_curve}
\end{equation}

\paragraph{Mechanical constraint on perturbations.}

In overdamped simple shear, force balance requires the total shear
stress to be spatially uniform,
\begin{equation}
    \partial_z\sigma=0.
\end{equation}
The imposed mean shear rate supplies the additional constraint
\begin{equation}
    \langle\dot\gamma\rangle=\overline{\dot\gamma},
    \qquad
    \langle\delta\dot\gamma\rangle=0.
\end{equation}
Here $\langle\cdot\rangle$ denotes the spatial average.

Consider a perturbation with zero spatial mean,
\begin{equation}
    \begin{pmatrix}
        \delta\alpha\\
        \delta\sigma_k\\
        \delta\dot\gamma
    \end{pmatrix}
    =
    \begin{pmatrix}
        \widehat\alpha\\
        \widehat\sigma_k\\
        \widehat{\dot\gamma}
    \end{pmatrix}
    \phi(z)e^{\omega t},
    \qquad
    \langle\phi\rangle=0.
\end{equation}
This perturbation allows different regions to shear at different
rates while maintaining the prescribed mean.

Define the differential viscosity of the non-bonded contribution, which we label $\sigma_0$
\begin{equation}
    H
    =
    \left.
    \frac{d\sigma_0}{d\dot\gamma}
    \right|_{\overline{\dot\gamma}}
    =
    Kn\overline{\dot\gamma}^{n-1}>0.
\end{equation}
The heterogeneous component of the total-stress perturbation must
vanish. Therefore,
\begin{equation}
    H\widehat{\dot\gamma}+\widehat\sigma_k=0,
    \qquad
    \widehat{\dot\gamma}
    =
    -\frac{\widehat\sigma_k}{H}.
    \label{eq:perturbed_force_balance}
\end{equation}

\paragraph{Linearized dynamics.}

Linearizing \eqref{eq:structure_local}
and \eqref{eq:stress_local} gives
\begin{align}
    \partial_t\delta\alpha
    &=
    -A\delta\alpha
    -b_\alpha\overline\alpha\delta\dot\gamma,
    \\
    \partial_t\delta\sigma_k
    &=
    G\overline{\dot\gamma}\delta\alpha
    -B\delta\sigma_k
    +(G\overline\alpha-b_\sigma\overline{\sigma_k})
      \delta\dot\gamma.
\end{align}
Using \eqref{eq:perturbed_force_balance} and the identity
\begin{equation}
    G\overline\alpha-b_\sigma\overline{\sigma_k}
    =
    \frac{G\overline\alpha k_-}{B},
\end{equation}
we obtain
\begin{equation}
    \omega
    \begin{pmatrix}
        \widehat\alpha\\
        \widehat\sigma_k
    \end{pmatrix}
    =
    \mathbf{J}
    \begin{pmatrix}
        \widehat\alpha\\
        \widehat\sigma_k
    \end{pmatrix},
    \qquad
    \mathbf{J}
    =
    \begin{pmatrix}
        -A & b_\alpha\overline\alpha/H\\
        G\overline{\dot\gamma} &
        -B-G\overline\alpha k_-/(BH)
    \end{pmatrix}.
    \label{eq:stability_matrix}
\end{equation}

Writing
\begin{equation}
    C=B+\frac{G\overline\alpha k_-}{BH},
\end{equation}
the two growth rates are
\begin{equation}
    \omega_\pm
    =
    -\frac{A+C}{2}
    \pm
    \frac{1}{2}
    \sqrt{
        (A-C)^2
        +\frac{4G b_\alpha\overline\alpha\overline{\dot\gamma}}{H}
    }.
    \label{eq:growth_rates}
\end{equation}
Since model parameters are positive, both eigenvalues are real and
$\operatorname{tr}\mathbf{J}=-(A+C)<0$.
Consequently, instability occurs when
$\det\mathbf{J}<0$, in which case exactly one eigenvalue is positive.
The instability is nonoscillatory.

\paragraph{Instability criterion.}

The determinant is
\begin{equation}
    \det\mathbf{J}
    =
    AB+\frac{AG\overline\alpha k_-}{BH}
    -\frac{G b_\alpha\overline\alpha\overline{\dot\gamma}}{H}.
\end{equation}
Differentiating the homogeneous flow curve yields
\begin{equation}
    \frac{d\overline\sigma}{d\overline{\dot\gamma}}
    =
    H+
    \frac{
        Gk_+
        \left[
            (k_++k_-)k_-
            -b_\alpha b_\sigma\overline{\dot\gamma}^2
        \right]
    }{A^2B^2}.
    \label{eq:flow_curve_slope}
\end{equation}
These expressions satisfy
\begin{equation}
    \det\mathbf{J}
    =
    \frac{AB}{H}
    \frac{d\overline\sigma}{d\overline{\dot\gamma}}.
\end{equation}
Since $A$, $B$, and $H$ are positive, the homogeneous state is
linearly unstable to heterogeneous perturbations precisely when
\begin{equation}
    \frac{d\overline\sigma}{d\overline{\dot\gamma}}<0.
    \label{eq:negative_slope_criterion}
\end{equation}
Equivalently, the unstable region satisfies
\begin{equation}
    Kn\overline{\dot\gamma}^{n-1}
    <
    \frac{
        Gk_+
        \left[
            b_\alpha b_\sigma\overline{\dot\gamma}^2
            -(k_++k_-)k_-
        \right]
    }{
        (k_++k_-+b_\alpha\overline{\dot\gamma})^2
        (k_-+b_\sigma\overline{\dot\gamma})^2
    }
    .
    \label{eq:dimensional_instability_criterion}
\end{equation}

\paragraph{Nondimensionalization.}

Recall the parameters
\begin{equation}
    p_0=\frac{k_+}{k_++k_-},
    \qquad
    \lambda_k=k_++k_-,
\end{equation}
so that $k_+=p_0\lambda_k$ and $k_-=(1-p_0)\lambda_k$.
Here $p_0$ is the zero-shear steady-state bond occupancy and
$\lambda_k^{-1}$ is its relaxation time. Since $K$ has dimensions
of stress times time to the power $n$, while $G$ has dimensions
of stress, we introduce the reference time
\begin{equation}
    \tau_0=\left(\frac{K}{G}\right)^{1/n}.
\end{equation}
For $\lambda_k>0$, we define the dimensionless kinetic and
shear rates as
\begin{equation}
    \Lambda_k=\lambda_k\tau_0,
    \qquad
    \widetilde{\dot\gamma}
    =\frac{\overline{\dot\gamma}}{\lambda_k}.
    \label{eq:dimensionless_rates}
\end{equation}
In these variables, \eqref{eq:dimensional_instability_criterion}
becomes
\begin{equation}
    n\Lambda_k^n\widetilde{\dot\gamma}^{\,n-1}
    <
    \frac{
        p_0
        \left[b_\alpha b_\sigma\widetilde{\dot\gamma}^{\,2}-(1-p_0)\right]
    }{
        (1+b_\alpha\widetilde{\dot\gamma})^2
        (1-p_0+b_\sigma\widetilde{\dot\gamma})^2
    }.
    \label{eq:dimensionless_instability_criterion}
\end{equation}
Thus $K$ and $G$ set the reference timescale, while the dimensionless
stability condition depends on $n$, $b_\alpha$, and $b_\sigma$.
The constant yield stress $\sigma_Y$ does not enter this condition.

\subsection{Phase boundary for the instability of homogeneous shear}
\label{sec:phase_boundary}

We determine the boundary in the $(p_0,\Lambda_k)$ plane separating
parameters that admit instability at some imposed shear rate from
those stable at every positive imposed shear rate. We again note that
$\lambda_k, b_\alpha,b_\sigma>0$, and $0<n<1$.

At fixed $p_0$ and $\Lambda_k$,
\eqref{eq:dimensionless_instability_criterion} can be written as
\begin{equation}
    \Lambda_k^n<\mathcal{F}(\widetilde{\dot\gamma};p_0),
    \label{eq:scaled_instability_criterion}
\end{equation}
where
\begin{equation}
    \mathcal{F}(\widetilde{\dot\gamma};p_0)
    =
    \frac{
        p_0 \widetilde{\dot\gamma}^{1-n}
        \left[b_\alpha b_\sigma \widetilde{\dot\gamma}^2-(1-p_0)\right]
    }{
        n(1+b_\alpha \widetilde{\dot\gamma})^2
        (1-p_0+b_\sigma \widetilde{\dot\gamma})^2
    }.
    \label{eq:instability_function}
\end{equation}

For $0<p_0<1$, $\mathcal{F}$ is positive only when
\begin{equation}
    \widetilde{\dot\gamma}>\widetilde{\dot\gamma}_0(p_0)
    =\sqrt{\frac{1-p_0}{b_\alpha b_\sigma}}.
\end{equation}
It vanishes at $\widetilde{\dot\gamma}=\widetilde{\dot\gamma}_0$ and approaches zero as $\widetilde{\dot\gamma}\to\infty$, so
it attains a positive maximum at a finite dimensionless shear rate.
Maximizing over $\widetilde{\dot\gamma}$ therefore gives the phase boundary
\begin{equation}
    \Lambda_{k,c}(p_0)
    =
    \left[
        \max_{\widetilde{\dot\gamma}>\widetilde{\dot\gamma}_0(p_0)}\mathcal{F}(\widetilde{\dot\gamma};p_0)
    \right]^{1/n}.
    \label{eq:kinetic_phase_boundary}
\end{equation}
For $\Lambda_k<\Lambda_{k,c}$, an interval of unstable imposed
shear rates exists. For $\Lambda_k>\Lambda_{k,c}$, the homogeneous
state is linearly stable at every positive imposed shear rate.
Equality corresponds to marginal stability. This condition can be solved numerically.

At $p_0=0$, the bond contribution vanishes and no instability
occurs. At $p_0=1$, we have the explicit expression
\begin{equation}
    \Lambda_{k,c}(p_0 = 1)
    =
    \left[
        \frac{b_\alpha^n}{4nb_\sigma}
        (1-n)^{1-n}(1+n)^{1+n}
    \right]^{1/n}.
\end{equation}

For all $p_0$, our derived phase boundary describes the loss of linear stability of homogeneous
flow. It does not, by itself, establish the existence or persistence
of a nonlinear banded state.

\end{document}